\documentclass{moriond}

\def\be{\begin{equation}}
\def\ee{\end{equation}}
\def\bea{\begin{eqnarray}}
\def\eea{\end{eqnarray}}

\newcommand{\Photo}{\includegraphics[height=35mm]{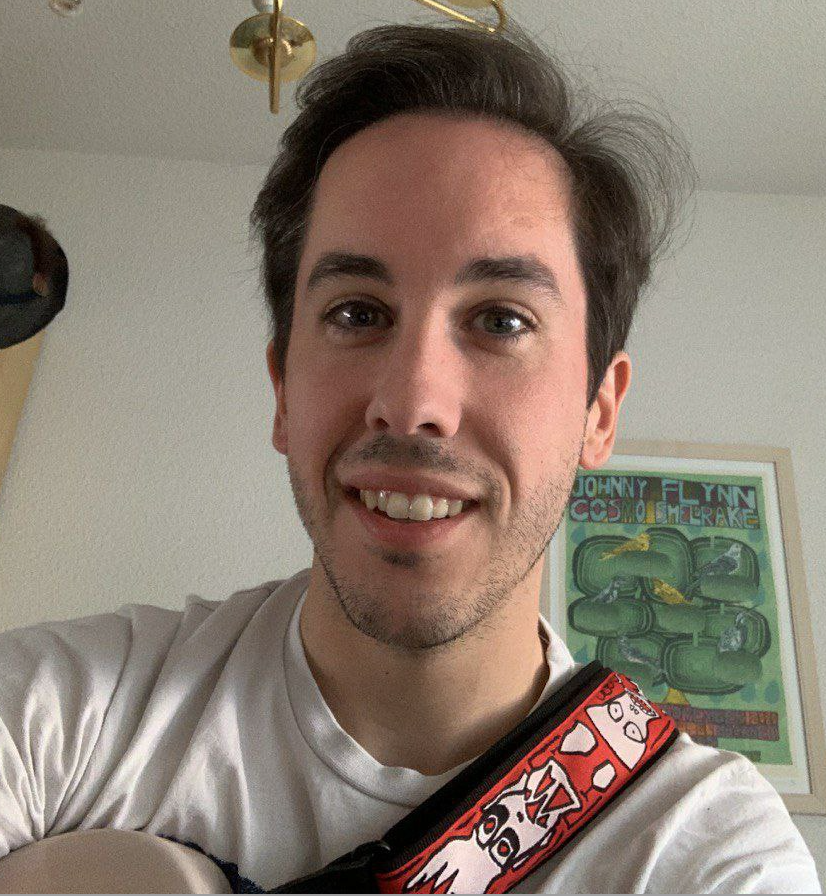}}

\begin{document}
\vspace*{4cm}
\title{$\nu$ Electroweak Baryogenesis}

\author{S.~ROSAURO-ALCARAZ}

\address{Departamento de F\'isica Te\'orica and Instituto de F\'{\i}sica Te\'orica, IFT-UAM/CSIC,\\
Universidad Aut\'onoma de Madrid, Cantoblanco, 28049, Madrid, Spain}

\maketitle\abstracts{
We investigate if the CP violation necessary for successful electroweak baryogenesis may be sourced by the neutrino Yukawa couplings. In particular, we consider an electroweak scale Seesaw realisation with sizeable Yukawas where the new neutrino singlets form (pseudo)-Dirac pairs. We find that flavour effects critically impact the final asymmetry obtained and that, taking them into account, the observed value may be obtained in some regions of the parameter space. This source of CP violation naturally avoids the strong constraints from electric dipole moments and links the origin of the baryon asymmetry of the Universe with the mechanism underlying neutrino masses. Interestingly, the mixing of the active and heavy neutrinos needs to be sizeable and could be probed at the LHC or future collider experiments.}

\section{Introduction}
The origin of the baryon asymmetry of the Universe (BAU) remains one of the most intriguing open questions of the Standard Model (SM). It has been measured~\cite{Aghanim:2018eyx} to be
\begin{equation}
     Y_B^{obs}\equiv \frac{n_b-n_{\bar{b}}}{s}\equiv \frac{n_B}{s}\simeq \left(8.59\pm 0.08\right)\times10^{-11},
    \label{Eq:YB_exp}
\end{equation}
where $n_b$ ($n_{\bar{b}}$) is the (anti)baryon number density and $s$ is the entropy density. To dynamically generate the BAU, the three Sakharov conditions~\cite{Sakharov:1967dj} need to be satisfied: baryon number violation, C and CP violation, and departure from thermal equilibrium. Although the SM in principle satisfies all three conditions, the amount of CP violation in the quark sector is not enough to generate the BAU~\cite{Gavela:1993ts} and the phase transition is rather a crossover~\cite{Kajantie:1996mn,Degrassi:2012ry} given the observed mass of the Higgs boson. 

Thus, a dynamical generation of the BAU requires physics beyond the SM (BSM). As a minimal option, an extension of the SM scalar sector could make Electroweak Baryogenesis~\cite{Shaposhnikov:1987tw} (EWBG) viable. In particular, new scalars could induce a strong first order phase transition~\cite{Dine:1990fj} at the EW scale and also contribute with new sources of CP violation. In this case, all the interesting physics would be around $\mathcal{O}(100)$~GeV, at reach of the Large Hadron Collider (LHC). However, new sources of CP violation typically induce electric dipole moments, which are very tightly  constrained~\cite{ACME_CP}, so EWBG models usually rely on some dark sector to avoid them. 

Given that the experimental evidence for neutrino masses from the observation of neutrino oscillations~\cite{Fukuda:1998mi} is also at odds with the SM, it represents another main window to new physics. It is therefore interesting to consider whether new sources of CP violation associated to the neutrino mass mechanism can generate the observed BAU. 

Here we will investigate the viability of EWBG in the context of low-scale Seesaw mechanisms~\cite{Mohapatra:1986aw,Mohapatra:1986bd,Bernabeu:1987gr,Malinsky:2005bi} in which neutrino masses are generated from a soft breaking of lepton number. The heavy neutrinos will thus be arranged in (pseudo-)Dirac pairs. A new scalar singlet, which can be responsible for the required strong first order phase transition, will also induce the Dirac mass of the heavy neutrinos. Thus, a CP asymmetry in the SM neutrinos, sourced by neutrino Yukawa couplings, may be produced through reflections and transmissions with the bubble wall. The imbalance between neutrinos and antineutrinos will then be converted into a baryon asymmetry through sphaleron processes in the unbroken phase. The generated net baryon number then enters the broken phase as the bubbles expand, where sphalerons are no longer efficient and baryon number is frozen out. 
\section{Neutrino mass generation and CP violation}
The SM is simply extended by three singlet Dirac neutrinos and a scalar singlet:
\begin{equation}
    \mathcal{L}=-\bar{L}_L \tilde{H} Y_{\nu} N_R -\bar{N}_L \phi Y_N N_R  + h.c. - V\left(\phi^* \phi,H^{\dagger}H\right),
    \label{Eq:Lagrangian}
\end{equation}
where $\phi$ is the singlet scalar and $H$ is the Higgs doublet, $L_L$ is the lepton doublet and $N_{R(L)}$ is the right(left)-handed component of the new Dirac neutrinos. $Y_{\nu}$ and $Y_N$ are $3\times3$ Yukawa matrices. We will remain agnostic as to the specifics of the lepton-number-violating contribution responsible for the observed neutrino masses, and in what follows, we will neglect these small perturbations on the underlying lepton-number-conserving structure. The last term in Eq.~(\ref{Eq:Lagrangian}) refers to the scalar potential, which couples the Higgs doublet to the singlet scalar and can induce the strong first order phase transition~\cite{Espinosa:1993bs}. 

After spontaneous symmetry breaking (SSB), both the SM Higgs field and the singlet scalar develop a vacuum expectation value (vev), $v_H$ and $v_{\phi}$, respectively, which generate Dirac mass terms for the neutrino states, $m_D\equiv v_H Y_{\nu}/\sqrt{2}$ and $M_N\equiv v_{\phi} Y_N$. The mixing between the heavy neutrinos and the active states is given by the ratio between the Dirac masses, $\theta\equiv m_D M_N^{-1}$, and can be sizeable. It is nonetheless experimentally bounded to be $Tr[\theta\theta^\dagger]\leq 0.007$ at $2 \sigma$~\cite{Fernandez-Martinez:2016lgt}. There is only one source of CP violation not suppressed by the generally smaller charged lepton Yukawas~\cite{Hernandez:1996bu,Santamaria:1993ah}, which is associated to the following basis invariant~\cite{Jarlskog:1985ht,Jenkins:2007ip}
\begin{equation}
    \delta_{CP}=M_1^2 M_2^2 M_3^2(M_1^2-M_2^2)(M_2^2-M_3^2)(M_3^2-M_1^2)Im\left[(\theta^{\dagger}\theta)_{12}(\theta^{\dagger}\theta)_{23}(\theta^{\dagger}\theta)_{31}\right].
    \label{Eq:Jarlskog_Inv}
\end{equation}

\subsection{Generation of a CP asymmetry}
\begin{figure}
\centerline{\includegraphics[width=0.4\textwidth]{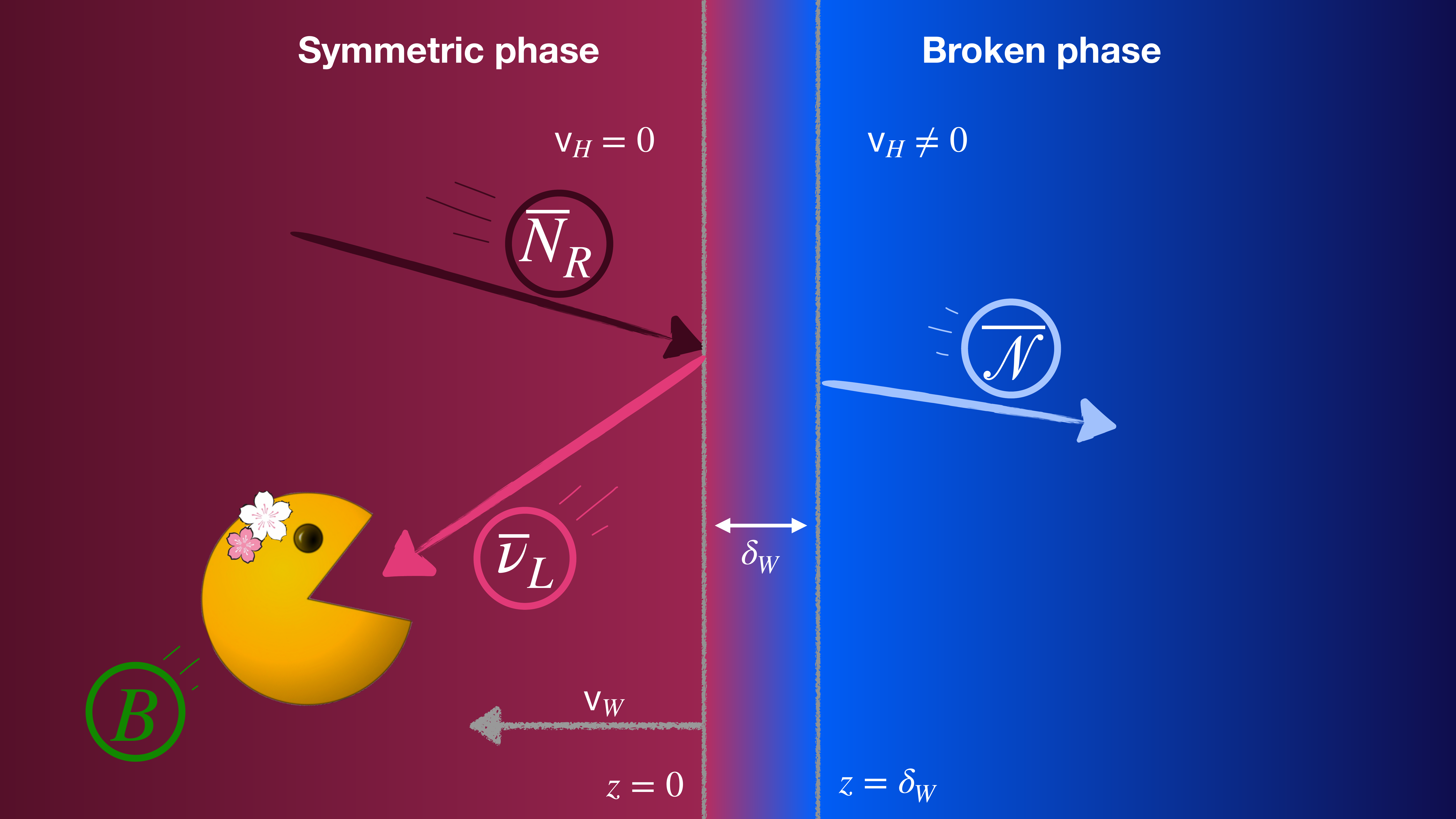}}
\caption{Sketch of the reflection of an incoming $\bar{N}_R$ off the bubble wall to a $\bar{\nu}_L$ and its subsequent conversion to baryons through sphaleron transitions.}
\label{fig:bubble}
\end{figure}
In the presence of the new scalar singlet, a strong first order phase transition is possible. Depending on the parameters of the scalar sector and its couplings to fermions, bubbles with walls of a given width $\delta_W$ will start nucleating at the temperature $T_c$ and expand at a velocity $v_W$. For details on the scalar potential such as the effect of the vev profiles, see Fern\'andez-Mart\'inez \textit{et al}~\cite{Fernandez-Martinez:2020szk}.

\begin{figure}
\begin{minipage}{0.5\linewidth}
\centerline{\includegraphics[width=0.8\linewidth]{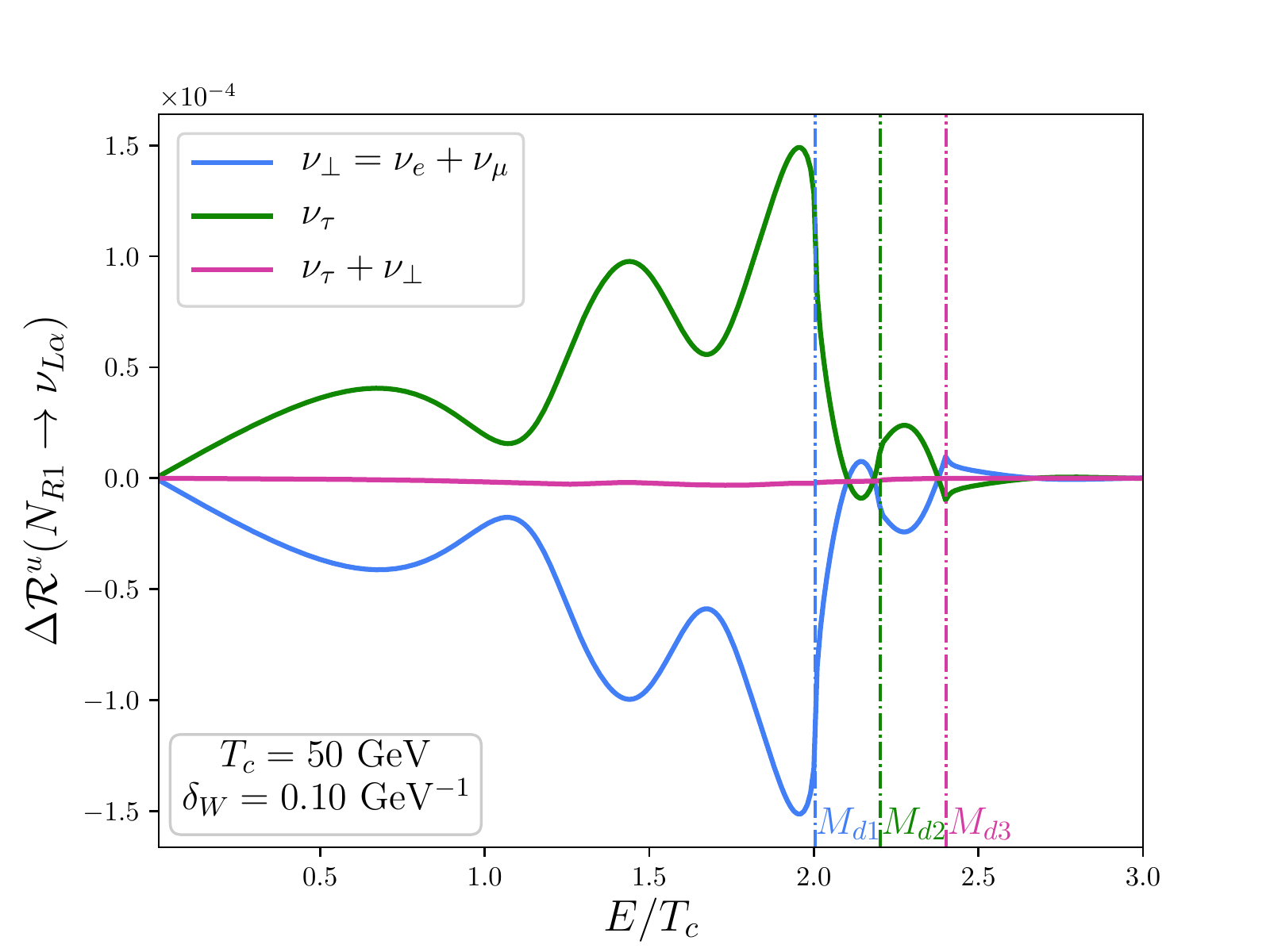}}
\end{minipage}
\hfill
\begin{minipage}{0.5\linewidth}
\centerline{\includegraphics[width=0.8\linewidth]{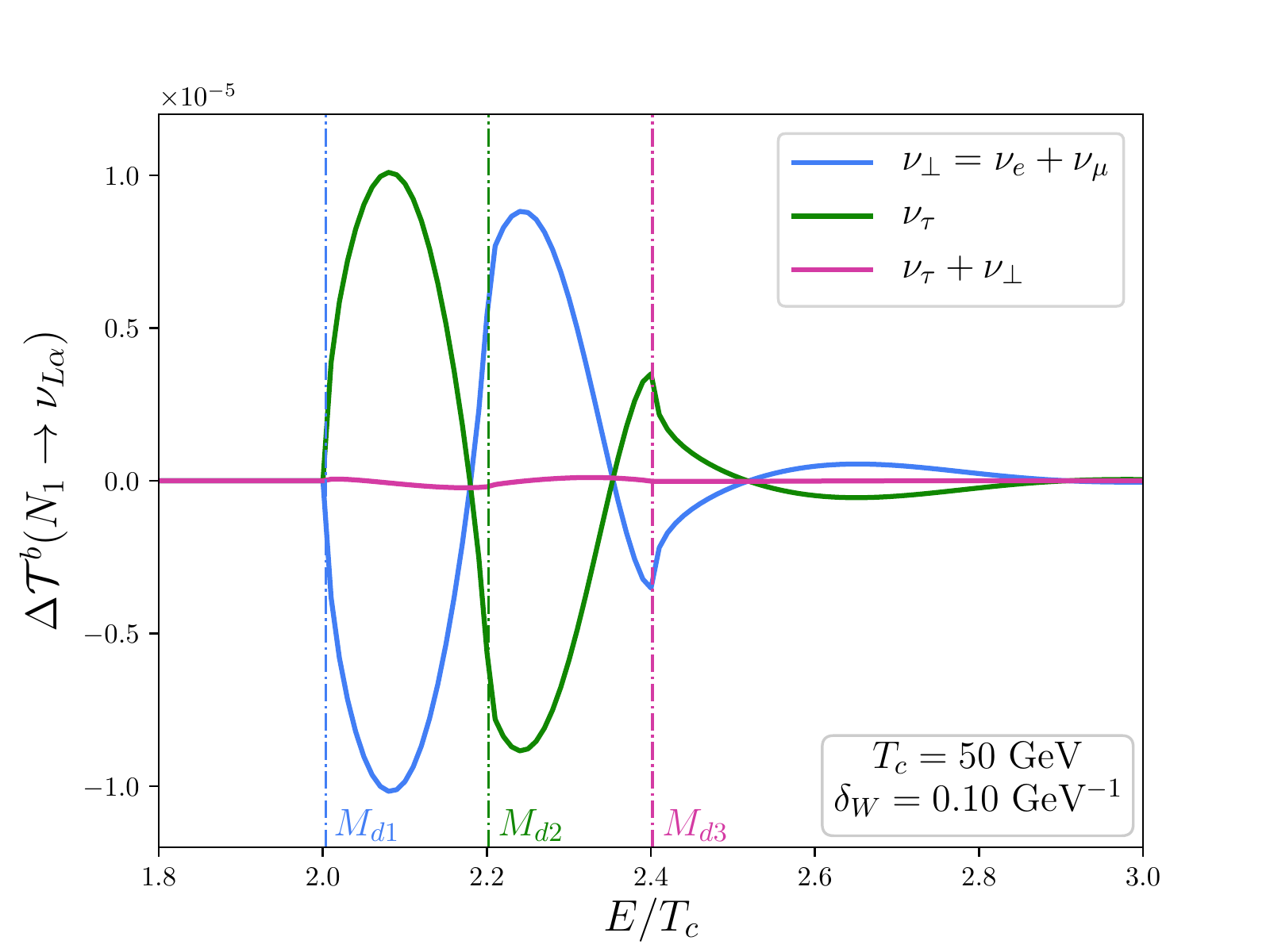}}
\end{minipage}
\caption{Flavoured CP asymmetries generated through reflections of $N_R$ from the symmetric phase (left panel) or transmission of heavy neutrinos $N$ from the broken to the symmetric phase (right panel) into each one of the neutrino flavours.}
\label{fig:ref_trans_flavour}
\end{figure}

Neutrinos travelling from the unbroken phase towards the bubble wall will be reflected as depicted in Fig.~\ref{fig:bubble}. In the presence of CP violation, the reflection rate for neutrinos and antineutrinos will be different, generating an asymmetry in $\nu_L$ (SM neutrinos) which will be converted into a baryon one through sphaleron transitions. The reflection and transmission asymmetries for each flavour, as computed in Fern\'andez-Mart\'inez \textit{et al}~\cite{Fernandez-Martinez:2020szk}, are shown in Fig.~\ref{fig:ref_trans_flavour}. 
We show in green the asymmetry stored in $\nu_{L\tau}$ and the sum of the remaining flavours, $\nu_{Le}$ and $\nu_{L\mu}$, in blue. As can be observed, their values are very similar and of opposite sign, such as the sum of the two, the total CP asymmetry (magenta line), is orders of magnitude smaller than the individual flavour asymmetries. This is due to a strong GIM-like cancellation appearing when summing over flavours~\cite{Fernandez-Martinez:2020szk}.
\section{Diffusion equations}
Once the CP asymmetries are generated, we need to solve a set of differential equations which follow the diffusion of each particle species. In the following we will study two cases: the simplest scenario where we only follow the total lepton and baryon asymmetries, dubbed as ``vanilla'' scenario, and another case where we include possible wash-out and flavour effects, the ``flavoured'' scenario.
\subsection{Vanilla scenario}
The minimal set of the diffusion equations we consider, following Hern\'andez \& Rius~\cite{Hernandez:1996bu}, is
\bea
        &D_B \partial_z^2n_B-v_W \partial_zn_B-3\Gamma_S\mathcal{H}(-z) n_B-\Gamma_S\mathcal{H}(-z) n_L = 0,\nonumber \\
        &D_L \partial_z^2n_L-v_W \partial_zn_L-\Gamma_S\mathcal{H}(-z) n_L-3\Gamma_S\mathcal{H}(-z) n_B = \xi_L j_{\nu}\partial_z\delta(z),
\label{Eq:Diff_Vanilla}
\eea
where we only follow the evolution of total baryon ($n_B$) and lepton number ($n_L$) asymmetries and their conversion through weak sphaleron processes. In Eq.~(\ref{Eq:Diff_Vanilla}), $D_{B(L)}$ is the diffusion constant for baryons (leptons) which we estimate following Joyce \textit{et al}~\cite{Joyce:1994zn}, $v_W$ is the wall velocity, $\Gamma_S = 9\kappa \alpha_W^5 T$ is the sphaleron rate with $\kappa\simeq 18$~\cite{DOnofrio:2014rug}, $\alpha_W$ the weak coupling constant, $\mathcal{H}$ the Heaviside function and we have neglected the bubble wall width. The CP current generated through reflections and transmissions of neutrinos, $j_{\nu}$, can be computed as in Fern\'andez-Mart\'inez \textit{et al}~\cite{Fernandez-Martinez:2020szk}. 
\begin{figure}
\begin{minipage}{0.5\linewidth}
\centerline{\includegraphics[width=0.8\linewidth]{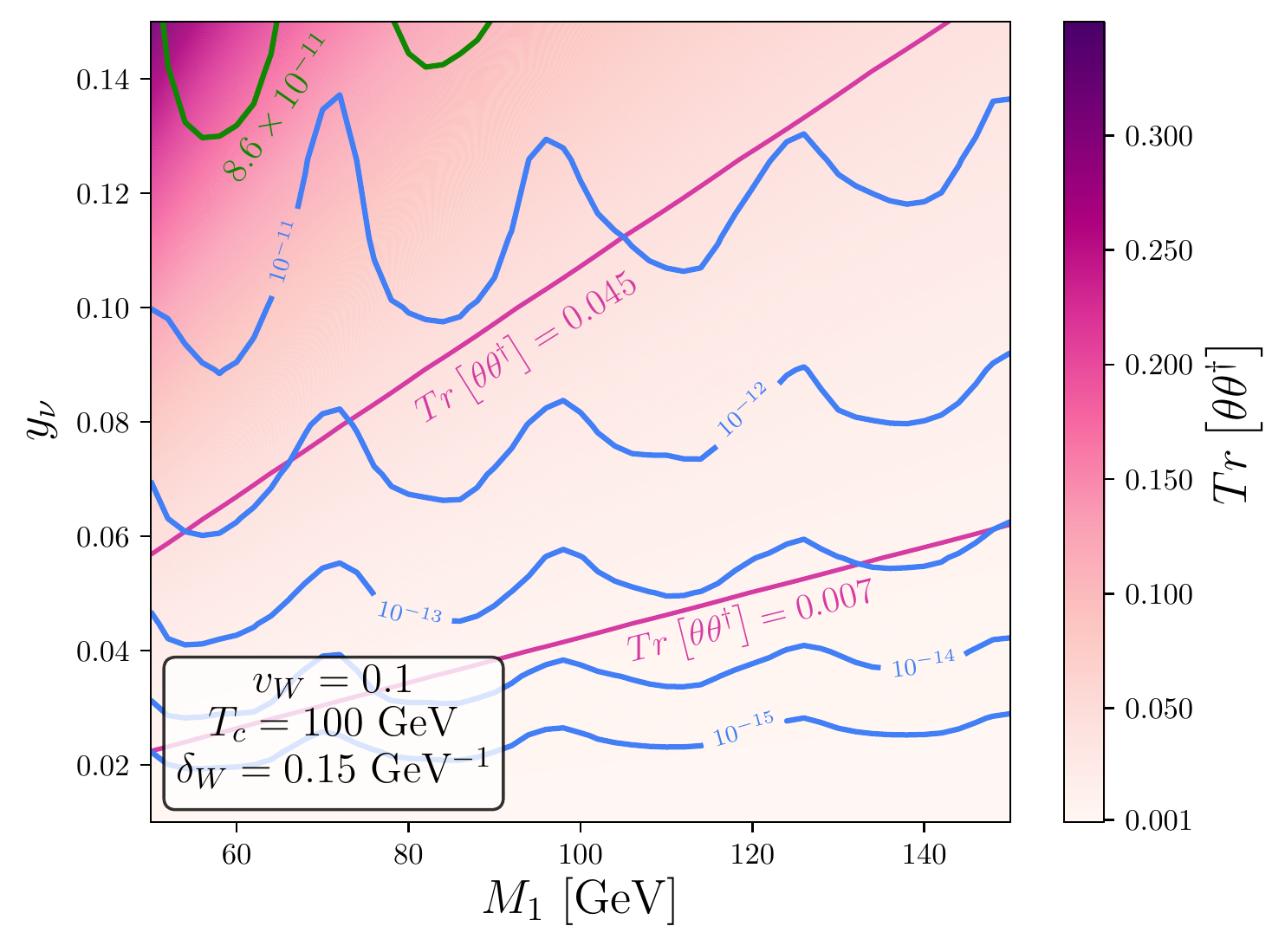}}
\end{minipage}
\hfill
\begin{minipage}{0.5\linewidth}
\centerline{\includegraphics[width=0.8\linewidth]{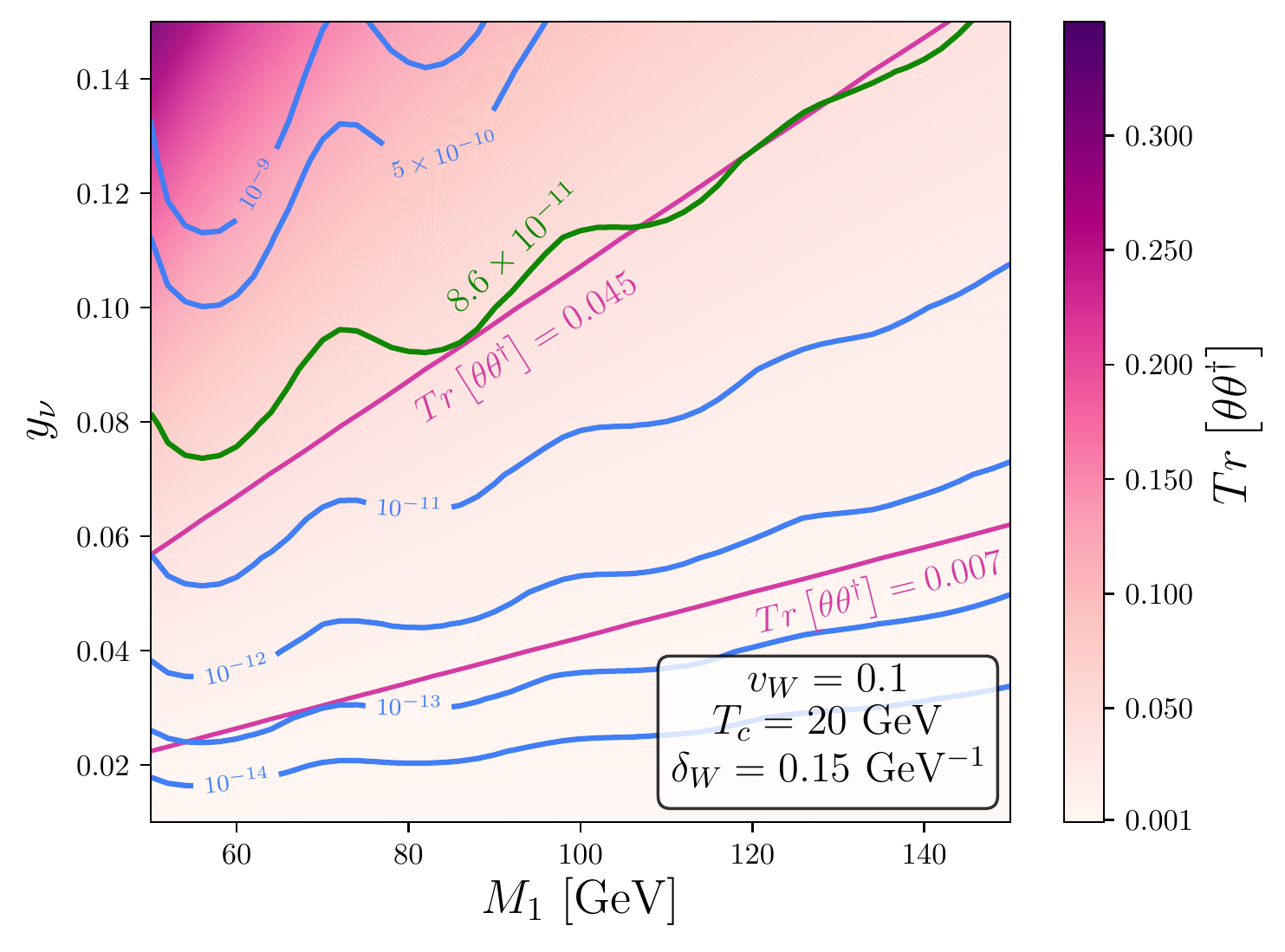}}
\end{minipage}
\caption{Contours of constant $Y_B$ as a function of the mass of the lightest heavy neutrino and the Yukawa coupling for the ``vanilla'' scenario.}
\label{fig:YB_Vanilla}
\end{figure}

We show in Fig.~\ref{fig:YB_Vanilla} contours of constant $Y_B$ as a function of the lightest heavy neutrino mass and the Yukawa coupling. The green contour corresponds to the observed baryon asymmetry while the two magenta contours corresponds to the bounds on $Tr[\theta\theta^{\dagger}]$ including or not the invisible width of the $Z$ boson. The left and right panels differ by $T_c$. From these plots it is clear that, in the ``vanilla'' scenario, we cannot explain the observed BAU unless the bounds on $\theta$ from the invisible width of the $Z$ boson can be avoided.
\subsection{Flavoured scenario}
\begin{figure}
\begin{minipage}{0.5\linewidth}
\centerline{\includegraphics[width=0.8\linewidth]{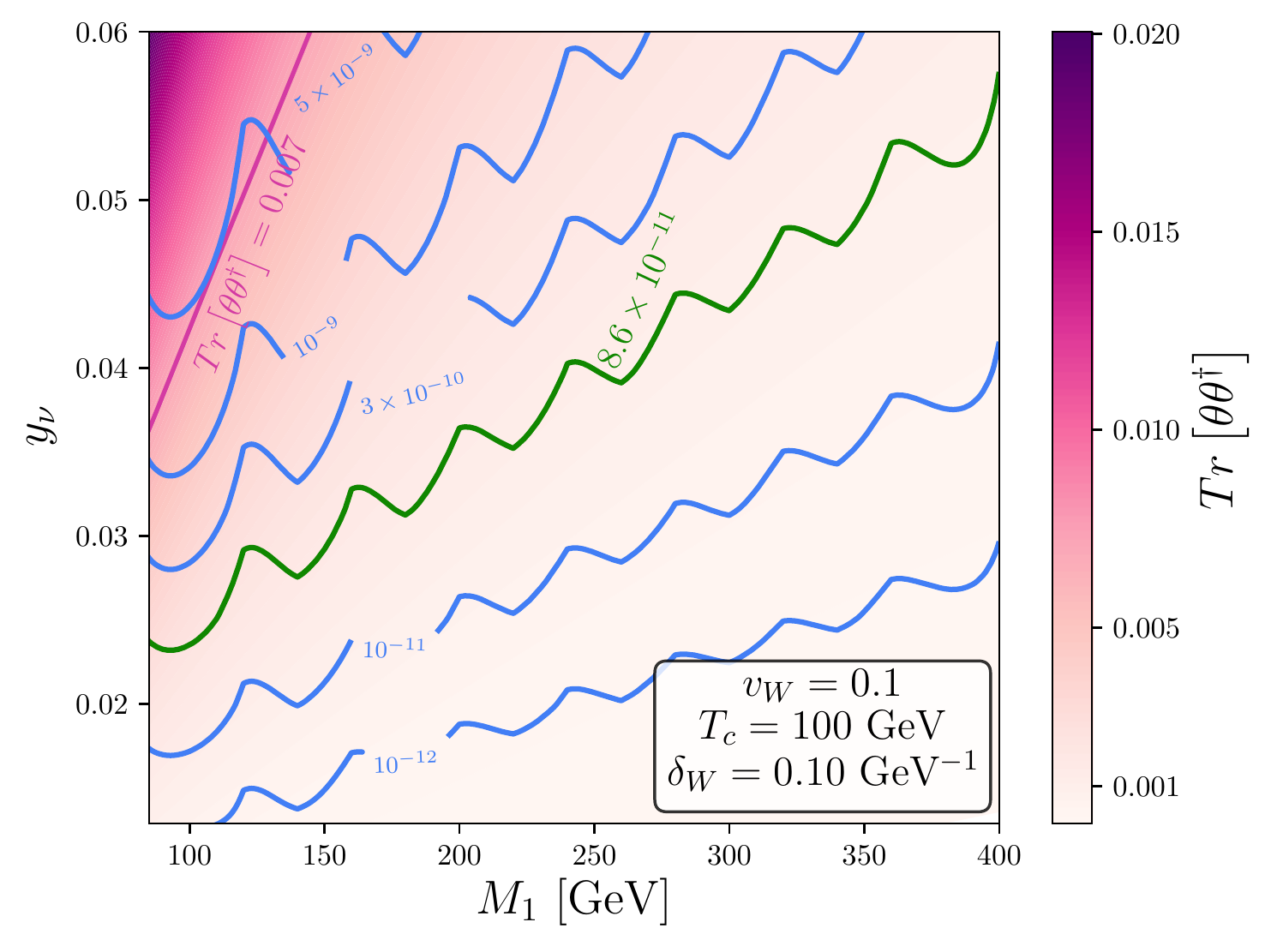}}
\end{minipage}
\hfill
\begin{minipage}{0.5\linewidth}
\centerline{\includegraphics[width=0.8\linewidth]{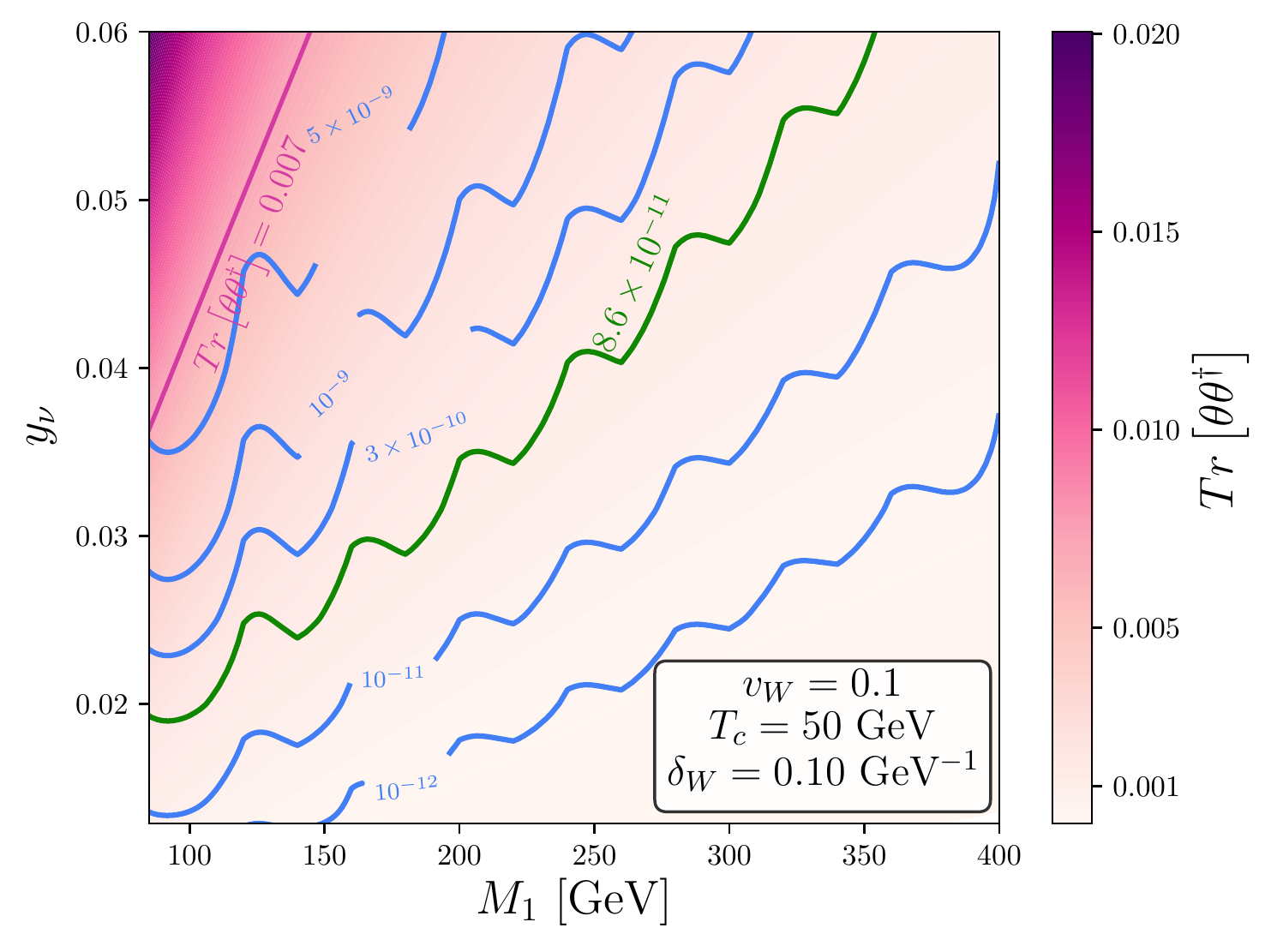}}
\end{minipage}
\caption{Contours of constant $Y_B$ as a function of the mass of the lightest heavy neutrino and the Yukawa coupling for the ``flavoured'' scenario.}
\label{fig:YB_Flavour}
\end{figure}

Given the strong GIM cancellation shown when summing over neutrino flavours (see Fig.~\ref{fig:ref_trans_flavour}), the introduction of possible wash-out effects from Yukawa interactions could prevent these suppression if these interaction rates are faster than the sphaleron rate. In particular, the wash-out from $\nu_L$ to $N_R$ is given by~\cite{Fernandez-Martinez:2020szk}
\begin{equation}
    \frac{\Gamma_{N_{Ri}\nu_{L\alpha}}}{T}\sim 0.0024|\theta_{\alpha i}|^2\frac{2M_{i}^2}{v_H^2},
    \label{Eq:Neutrino_Yukawa_rate}
\end{equation}
which is larger than the sphaleron rate for $M_i>200$~GeV given the bounds on $\theta$. Thus, this wash-out term can play a crucial role in breaking the GIM cancellation when following the individual flavour asymmetries. Additionally, we will have to follow as well the asymmetries stored in each $N_R$, which interact much less with the plasma, diffusing much more than $\nu_L$ into the symmetric phase, and potentially contributing to enlarging the final baryon asymmetry. As can be seen in Fig.~\ref{fig:YB_Flavour} and in more detail in Fern\'andez-Mart\'inez \textit{et al}~\cite{Fernandez-Martinez:2020szk}, this is indeed the case, and we can successfully explain the BAU respecting all experimental constraints.

\begin{figure}
\begin{minipage}{0.5\linewidth}
\centerline{\includegraphics[width=0.8\linewidth]{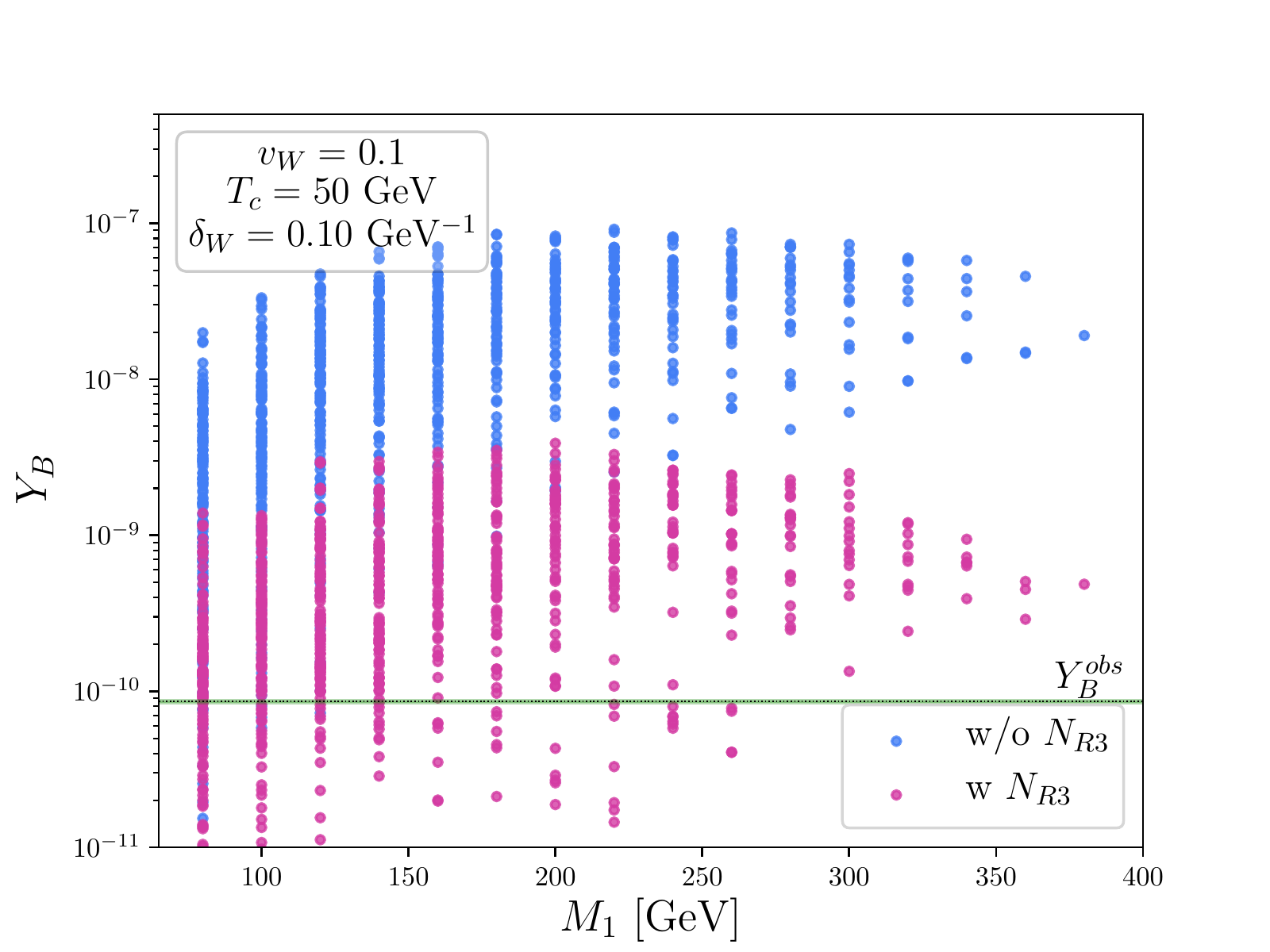}}
\end{minipage}
\hfill
\begin{minipage}{0.5\linewidth}
\centerline{\includegraphics[width=0.8\linewidth]{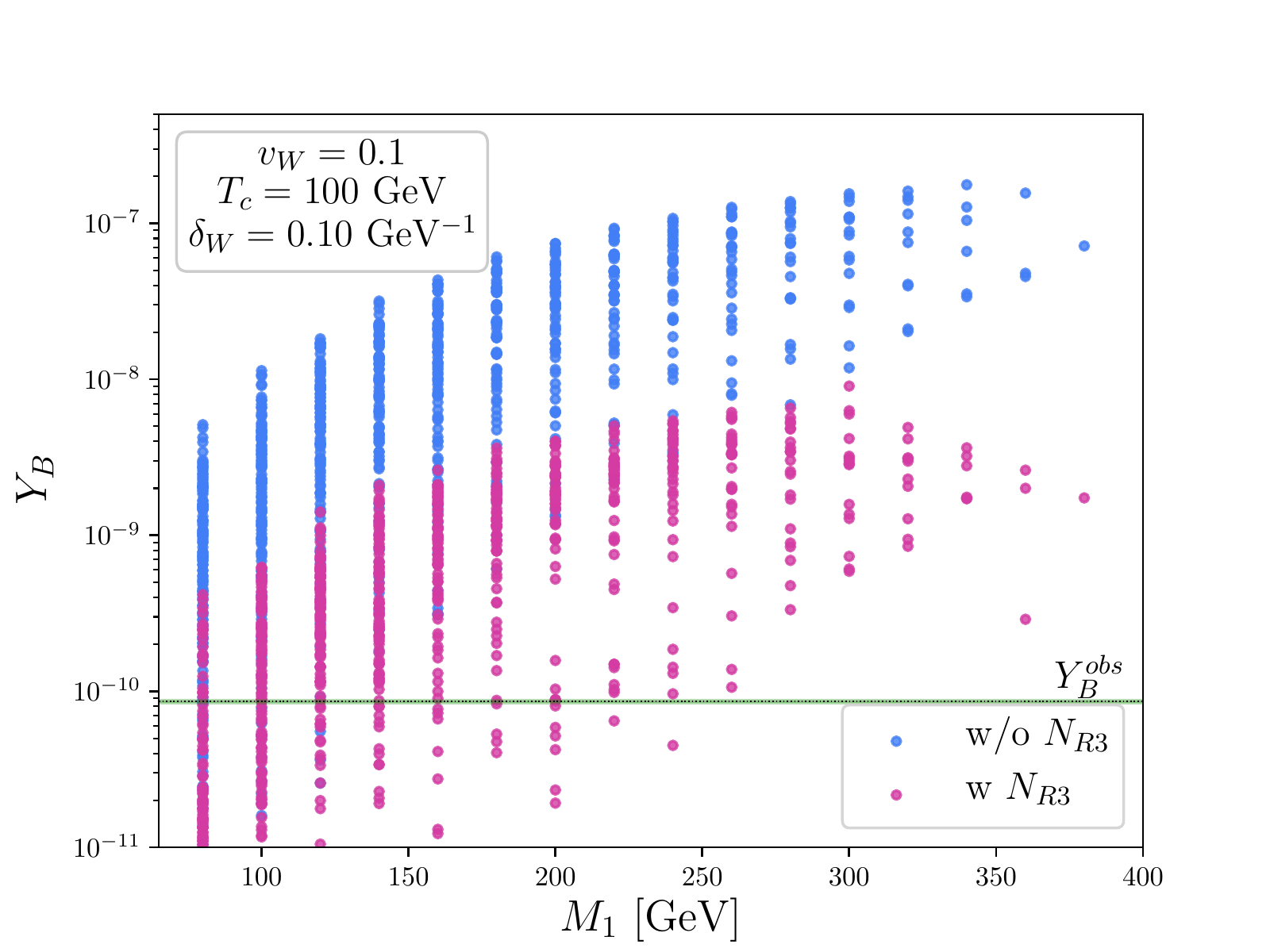}}
\end{minipage}
\caption{Values of $Y_B$ as a function of the masses for two different $T_c$. The blue and magenta dots correspond to neglecting or overestimating the effect of $N_{R3}$.}
\label{fig:YB_Mass}
\end{figure}

This is further confirmed in Fig.~\ref{fig:YB_Mass} where we show the values of $Y_B$ obtained by scanning over the masses of the heavy neutrinos. The blue points correspond to neglecting the effect of the heaviest neutrino, $N_{R3}$, while the magenta ones to overestimating its impact on the final $Y_B$. The actual contribution of $N_{R3}$ with its corresponding diffusion constant would yield a result lying between the two limits corresponding to the magenta and blue points. In both panels one can see that we can explain the observed BAU (green band). Given that we are here saturating the bounds on $\theta$, but the BAU is proportional to the sixth power of $\theta$, an improvement of a factor of $\sim2$ in $\theta$ would reduce the final $Y_B$ by a factor $\sim 100$, eventually allowing to probe the entire parameter space.
\section{Conclusions}
We have studied the possibility that the mechanism responsible for neutrino masses also helps in the production of the BAU within the context of electroweak baryogenesis. Low scale realisations of the Seesaw mechanism not only naturally explain the origin of tiny neutrino masses through an approximate lepton number symmetry, but are also testable since the heavy neutrinos can be at the electroweak scale with a sizeable mixing with their active partners. It is therefore tantalising to explore the role of these new states and sources of CP violation at the electroweak scale in electroweak baryogenesis. Furthermore, the neutrino sector naturally avoids the problematic EDM constraints. 

We have investigated two scenarios where the baryon asymmetry is generated from the CP violation stemming from the neutrino Yukawa couplings. In the simplest case, neglecting the flavour-dependent Yukawa rates, we find it is not possible to explain the observed BAU unless present bounds on heavy-active neutrino mixing can be avoided. This idea was first studied by Her\'andez \& Rius~\cite{Hernandez:1996bu}, which was the starting point of our analysis that we expanded in several aspects, such as the inclusion of flavour-dependent wash-out effects. More interestingly, when the flavour-dependent wash-out rates are included, the observed BAU can be successfully explained within present constraints. In any event, the required mixing is always large and these scenarios could be testable by future collider searches. 

\section*{Acknowledgements}
SRA warmly thanks his collaborators E.~Fern\'andez-Mart\'inez, J.~L\'opez-Pav\'on and T.~Ota, without whom this work could not have been possible, and acknowledges the support of the Spanish Agencia Estatal de Investigaci\'on and the EU ``Fondo Europeo de Desarrollo Regional'' (FEDER) through the projects PID2019- 108892RB-I00/AEI/10.13039/501100011033 and FPA2016-78645-P as well as the ``IFT Centro de Excelencia Severo Ochoa SEV-2016-0597''.
\section*{References}
\bibliography{Baryogenesis_bib}
\end{document}